\documentclass[twocolumn,prb,superscriptaddress]{revtex4}
\newcommand{\be}{\begin{equation}}
\newcommand{\ee}{\end{equation}}

\newcommand{\bfr}{ {\bf r} }
\newcommand{\bfz}{ {\bf z} }
\newcommand{\bfk}{ {\bf k} }
\newcommand{\bfG}{ {\bf G} }

\newcommand{\bfq}{ {\bf q} }
\newcommand{\bfp}{ {\bf p} }
\newcommand{\V}{ {V} }
\newcommand{\calH}{ {\hat {\cal H}} }
\usepackage{graphicx}
\usepackage{dcolumn}
\usepackage{bm}
\usepackage{psfig}

\begin{document}
\title{Backflow Correlations for the Electron Gas and Metallic Hydrogen}
\author{M. Holzmann}
\affiliation{Laboratoire de Physique Th\'eorique des Liquides, UMR
7600 of CNRS, Universit\'e P. et M. Curie, boite 121, 4 Place
Jussieu, F-75252 Paris, France} \affiliation{University of Illinois at
Urbana-Champaign, Urbana, IL 61801, USA}
\author{D. M. Ceperley}
\affiliation{University of Illinois at Urbana-Champaign, Urbana,
IL 61801, USA} \affiliation{CECAM, ENS,46 allee d'Italie, F-69364
Lyon, France}
\author{C. Pierleoni}
\affiliation{CECAM, ENS,46 allee d'Italie, F-69364 Lyon, France}
\affiliation{INFM and Department of Physics, University of
L'Aquila, Via Vetoio, I-67010 L'Aquila, Italy}
\author{K. Esler}
\affiliation{University of Illinois at Urbana-Champaign, Urbana,
IL 61801, USA}


\begin{abstract}
We justify and evaluate backflow-threebody wavefunctions for a two
component system of electrons and protons. Based on the
generalized Feynman-Kacs formula, many-body perturbation theory,
and band structure calculations, we analyze the use and the
analytical form of the backflow function from different points of
view. The resulting wavefunctions are used in Variational and
Diffusion Monte Carlo calculations of the electron gas and of
solid and liquid metallic hydrogen. For the electron gas, the
purely analytic backflow and three-body form gives lower energies
than those of previous calculations. For bcc hydrogen, analytical
and optimized backflow-threebody wavefunctions lead to energies
nearly as low as those from using LDA orbitals in the trial
wavefunction. However, compared to wavefunctions constructed from
density functional solutions, backflow wavefunctions  have the
advantage of only few parameters to estimate, the ability to
include easily and accurately electron-electron correlations, and
that they can be directly generalized from the crystal to a
disordered liquid of protons.

\end{abstract}

\maketitle



\section{Introduction}

This paper concerns the form of the ground state wavefunction of
metallic hydrogen at high enough density so that all the hydrogen
molecules are dissociated and the electrons are delocalized.
Neglecting possible quantum effects on the protonic motion, the
many-body wavefunction can be regarded as the ground state of an
electron gas under the influence of an external potential due to
the actual positions of the protons. Quantum Monte Carlo (QMC)
techniques are currently one of the most powerful methods to
calculate accurately the properties of such a many-body quantum
system\cite{RMPQMC}. However, since ground state QMC is based on
trial wavefunctions, QMC typically demands compact {\em and}
accurate descriptions of the ground state wavefunction. In this
paper we review different approaches to obtain and improve trial
wavefunctions, compare the qualities of the resulting many-body
wavefunctions with previous QMC calculations for the electron gas
and metallic crystal hydrogen, and present first results using
these wavefunctions for liquid metallic hydrogen.

Most of the work within QMC has been done using a pair product
(PP) (or Slater-Jastrow) wavefunction: a Slater determinant of
single electron spin orbits times a product of pair electron
(Jastrow) factors. Notwithstanding certain deficiencies such as a
lack of direct spin coupling, this wavefunction has proven to be
quite accurate, in particular within fixed-node Diffusion Monte
Carlo\cite{RMPQMC} (DMC). The first calculation on many-body
hydrogen\cite{dmc81} used an even simpler form of this
wavefunction; the single electron orbits were taken to be free
electron plane waves. We refer to this as the SJ-PW trial
function. Later, Natoli \cite{natoli93,natoli95a} found that
determinants using these orbitals are inaccurate by 0.05eV/atom
within the fixed-node DMC calculations at the density
corresponding to the transition between molecular and metallic
hydrogen ($r_s=1.31$). Hence, more accurate orbitals, computed
from either density functional (LDA) or Hartree-Fock (HF)
calculations, are required. Because these orbitals are calculated
assuming fixed ionic positions, inclusion of ionic motions, such
as those from the zero point motion of the ions in the crystal, is
difficult.

Recently, there have been new attempts\cite{dewing01,dmc03} to
calculate properties of disordered systems such as liquid hydrogen
within QMC. In the Coupled Electron Ion Monte Carlo (CEIMC)
method\cite{dewing01} the protons are moved based on the results
of a QMC calculation of the electronic energy. This approach
requires accurate trial functions that can be obtained quickly as
the ionic positions are changed; methods involving the solution of
mean field equations such as LDA and HF, or even optimizing a
parameterized trial function, can greatly slow down the overall
performance of the CEIMC simulation\cite{dewing01}. Further,
combining the orbitals obtained from LDA or HF with a pair
correlation (Jastrow) factor to improve the accuracy is not
straightforward; substantial modification of the orbitals might be
necessary requiring a reoptimization of the orbitals and
correlation factor, in principle, at each ionic
position\cite{fahy}. This optimization step creates a bottleneck
to coupling the QMC calculations with the ionic Monte Carlo.

One could consider obtaining the trial wavefunction from other
variational approaches like Fermi-hypernetted (FHNC) chain or
correlated basis functions (CBF) methods\cite{krotscheck}  which
would not have the problems of optimization. However, in these
approaches based on explicit integration, one is in general
limited in the form of the trial function by the ease performing
the integration, and these are typically much more time-consuming than
LDA calculations.

One of the biggest advantages of the QMC approach is that one can
use an arbitrary wavefunction without changing the algorithm in an
essential way. Fast algorithms will result if one can find concise
and accurate forms. In this paper, instead of using one-body
orbitals from mean field theory or integral equations, we propose
to use trial functions which depend explicitly and continuously on
the ionic variables. Such wavefunctions do not have to be
reoptimized for movements of the ions, are easy to implement, and
accurate for disordered systems. Calculation of ionic forces is
simplified since the derivative of the trial function with respect
to ionic configurations is a straightforward application of the
chain rule. These trial functions are a generalization of the
backflow three-body wavefunctions used very successfully in highly
correlated homogeneous quantum liquids: liquid $^3$He and the
electron gas. There, backflow trial functions show much
improvement over the pair product getting approximately 75\% of
the energy missing at the PP level and even more when done with
the fixed-node method.

Backflow wavefunctions were developed by Feynman and Cohen
\cite{feynman} for a single $^3$He impurity in liquid $^4$He when
it was recognized that without backflow, the mass of the impurity
was equal to the bare mass. Pandharipande and Itoh \cite{panda73}
showed that the backflow arises from the momentum dependence of
the correlation between the impurity and the liquid. The backflow
wavefunction was then extended to bulk liquid $^3$He
\cite{schmidt79,manousakis} using an integral equation method to
evaluate expectation values. The first use of backflow in QMC was
by Lee et al.\cite{lee81} and others \cite{schmidt81,panoff89}
with calculations on liquid $^3$He. Moroni et al.\cite{moroni95}
further optimized the trial function within liquid $^3$He. Kwon et
al.\cite{kwon93,kwon98} used backflow functions for the electron
gas in both 2D and 3D, obtaining significantly lower energies and
improved excitation energies. Vitiello et al.\cite{vitiello97}
discuss an equivalence of backflow and spin-dependent
correlations, an aspect we will not further consider in this
paper.

Using different approaches, we generalize the backflow three-body
wavefunction to a two component system of electrons and protons
and derive approximate expressions for the correlated trial
function. We first present an argument based on the generalized
Feynman-Kacs formula which shows that backflow is the next order
improvement beyond the pair product (PP) wave function. Using
perturbation theory, we then discuss general features of the
backflow functions and obtain explicit expressions for the
homogeneous electron gas and for the electron-proton plasma. A
similar analysis using the Bohm-Pines method has been recently
performed by Gaudoin et al.\cite{gaudoin}, however, without going
beyond the Slater-Jastrow wavefunction. Studying the problem of a
single electron in the potential generated by a simple cubic
lattice of protons, we show that the exact one electron
wavefunction can be approximately rewritten by a backflow
function. Finally, we optimize numerically simple functional forms
for the backflow functions in the full many-body problem by
variational Monte Carlo. We compare the quality of the
wavefunctions stemming from these different approaches for the
electron gas and for liquid and crystal hydrogen at the level of
variational and diffusion Monte Carlo.

In the following we consider the non-relativistic Hamiltonian of
$N$ protons and $N$ electrons: \be{\hat H}=-\sum_i \lambda_i
\nabla_i^2 +\sum_{i<j} \frac{e_ie_j}{r_{ij}}\ee where
$\lambda_i=\hbar^2/(2m_i)$, $i=1,\dots,2N$ and $m_i$ and $e_i$ are
the electron or proton mass and charge. The Fermi wavevector is
$k_F$. Numerical results are given in atomic units where
$\lambda_e=1/2$ and $\lambda_p=0$ for classical protons, $|e_i|=
m_i=1$. The electron density $n=N/V$ is quoted in terms
$r_s=a/a_0$, where $a=(4 \pi n/3)^{-1/3}$ and $a_0=\hbar^2/m_e e^2$
is the Bohr
radius. Energies of the QMC calculations are given in Rydbergs per
electron.

\section{The Feynman-Kacs approach to improving the wavefunction}

The Feynman-Kacs formula expresses the exact wavefunction in terms
of average over Brownian paths. We now review how it can be
generalized to random walks with ``drift''.

We define the ``importance-sampled'' Green's function as ${\hat
G}_t = \psi \exp(-t {\hat H}) \psi^{-1}$ in operator notation
where $\psi$ is an unsymmetrical trial function. ${\hat G}_t$
acting on a function has the effect of enhancing the component of
lower energy states. Then the lowest energy (exact) fermi
wavefunction $\phi_F (R) $ is given by: \be \phi_F (R) \propto
\hat{{\cal A}} \psi (R) \lim_{t\rightarrow \infty} \int dR'
\langle R' | {\hat G}_t | R \rangle \ee assuming only that the
trial function has a non-zero overlap with $\phi_F$ and that
$\phi_F$ is non-degenerate. Here $\hat{{\cal A}}$ is a projection
operator for fermion symmetry defined as\be \hat{{\cal A}} f(R) =
\frac{1}{N!}\sum_P (-1)^P f(P R)\label{proja} \ee and $R=\{\bfr_1,
\bfr_2, \ldots \}$ is a point in configuration space. The electron
spin is treated by restricting the permutation in
Eq.~(\ref{proja}) to be exclusively within spin up or spin down
electrons.

Following the derivation in Diffusion Monte Carlo\cite{reynolds},
the Green's function can be split into diffusion, drift and
branching processes. To show this, the master equation for the
Green's function is written: \be -\frac{d\hat{G}}{dt} = \psi
\hat{H}\psi^{-1} \hat{G} = [ -\sum_i \lambda_i \nabla_i ( \nabla_i
+ 2 \nabla_i \ln \psi ) + E(R) ] \hat{G} .\ee The local energy
$E$, defined as $E(R)= \psi^{-1} {\hat H} \psi $, is the residual
error of the trial function, and becomes a constant function as
$\psi$ approaches an exact eigenvalue. Trotter's formula applies
to the above master equation, allowing us to split up the
evolution into the first two terms describing a stochastic
process, and the final term which is a branching or ``weighting''
process. Thus we have the generalized Feynman-Kacs formula : \be
\phi_F (R) \propto {\cal A} \psi (R) \left< \exp \left[
-\int_0^{\infty} dt E (R(t) ) \right] \right> \ee where the
brackets imply averaging over all drifting random walks $R(t)$
beginning at a point $R$. The above relation is exact for any real
trial function. For trial functions having an imaginary component
of $\nabla \ln \psi$, the formalism goes through, however, the
Green's function is no longer real and positive and therefore
cannot be treated as a probability. Other methods are more
appropriate. For the moment we will ignore this case.

To make further analytical progress, we take the average into the
exponent. For any stochastic process, one can write the average of
the exponent as the exponential of the cumulant expansion, the
first two terms of which are: \be \phi_F (R) \propto {\cal A} \psi
(R) \exp( - \langle\langle E \rangle\rangle + (1/2) \langle\langle
\delta E^2 \rangle\rangle \ldots ]. \ee The double brackets are
defined as $\langle\langle E \rangle\rangle= \langle
\int_0^{\infty} dt E(R(t))\rangle$ with walks $R(t)$ generated
from the drift and diffusion starting at a point $R$.  We truncate
the cumulant expansion after the first term. We then have an
approximate method of improving the trial function. \be
\psi^{(n+1)}= \psi^{(n)} e^{-\langle\langle
E^{(n)}\rangle\rangle_n } \label{psi_iter}\ee  with the subscript
indicating that the drift is given by $\nabla \ln \psi^{(n)}$. If
we split the log of the trial function into its real and imaginary
parts $\psi^{(n)}= \exp(-U^{(n)} + i S^{(n)} )$ with $U$ and $S$
real, we are led to the following equations for a single
iteration:
\begin{eqnarray}
 U^{(n+1)}& = &U^{(n)}  \nonumber \\
 +  \langle\langle &V&+\sum_i\lambda_i[ \nabla_i^2
U^{(n)}-(\nabla_i U^{(n)})^2+(\nabla_i S^{(n)})^2\rangle\rangle_n
\nonumber\\
\end{eqnarray}
\be
S^{(n+1)}= S^{(n)}+ \langle\langle \sum_i \lambda_i [ \nabla_i^2
S^{(n)}-2\nabla_i U^{(n)}\nabla_i S^{(n)} ] \rangle\rangle_n .
\label{usiter}\ee Here $V(R)$ is the total potential energy.

Specializing to the case of a fermi liquid, we take as an
initial wavefunction $U^{(0)} = 0$ and $S^{(0)} = \sum_i
\bfk_i \cdot \bfr_i $, {\it i.e.} singly occupied free particle states.
(The usual spin functions are assumed but not explicitly written.)
Note that this function is an unsymmetrical trial function, with a
non-zero overlap with a fermion state as long as all the
$\bfk_i$'s are distinct. When the wavefunction is antisymmetrized,
one gets a determinant of plane waves. However, the
antisymmetrization will be done only once, {\it after} the trial
function has gone through several iterations of Eq.
(\ref{psi_iter}). This way simplify the procedure, since the local
energy of the unsymmetric trial function is much simpler than that
of an antisymmetric trial function. Note that in Eq. (5) both the
antisymmetrization and the averaging are linear operators and so
can be interchanged.

After the first iteration, the wavefunction will have the form:
\be U^{(1)} = \langle\langle V(R) \rangle\rangle_0 \equiv U(R)\ee
\be S^{(1)}=S^{(0)}.\ee In the above equation and the following
discussion we drop, without mention, constant normalization terms.
If $V(R)= \sum_{i<j} v(r_{ij})$ is a pair potential, with a
Fourier transform $v_k$, the averaging can be carried out
analytically with a result that $U^{(1)}$ will also be a pair
potential and will have a Fourier transform given by $v_k/(\lambda
k^2)$ where $\lambda = \lambda_i+\lambda_j$. For a Coulomb
potential the real-space correlation (Jastrow) function will then
have the form: $u_C(r)=- e^2 r/2\lambda$.

\begin{figure}
\centerline{\psfig{figure=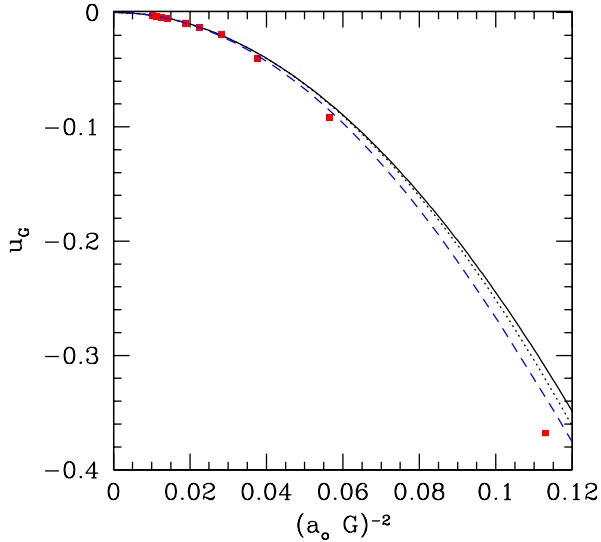,width=10cm,angle=0}}
\caption{The electron-proton Jastrow factor $u_G^{ep}$ versus
$(a_0G)^{-2}$ from band calculations of solid cubic hydrogen at
$r_s=1.31$ (squares), Eq. (53). The rightmost square is the first
reciprocal lattice vector. This is compared with the RPA (Gaskell)
form (solid line) Eq. (59) and cumulant form (dotted line) $1/G^4$
and the improved analytic form (Eq. 71) (dashed line). }
\label{jastrow}
\end{figure}

Hence, the form of the first order wavefunction is of the SJ-PW or
Slater-Jastrow form, with free particle orbitals. The pair term
will be denoted by $U$ with $U=\sum_{i<j} u(r_{ij} ) $. Typically
the form of $u$ is derived from a variational principle, chosen so
such that either the total energy or variance is minimized. This
will, of course, give a lower energy than the cumulant form
derived above. The above derivation does give the correct cusp
condition (the limit of $u$ at large $k$ or small $r$). However,
it does not give the long wavelength limit correctly because of
the neglect of the higher cumulants. Gaskell\cite{gaskell}
proposed an analytic form based on the Random Phase Approximation
(RPA) without any parameters. It was found\cite{dmc78} for the
homogeneous electron gas that the RPA form does, as well as, or
better than simple assumed forms with parameters. Figure
\ref{jastrow} shows a comparison of these correlation functions.

Note that the cumulant approximation will not exist if the Fourier
transform of the potential does not exist. Two examples of such
potentials are the hard sphere and Lennard-Jones interactions.
However, for the short range part of a soft potential which does
have a Fourier transform such as the Yukawa potential, the
cumulant approximation works quite well (see remarks concerning
the situation at finite temperature in Ref.~\cite{rmpi}).

We now perform the next iteration of this procedure. To minimize
the fluctuations in the local energy so that the cumulant
approximation will be more accurate, we assume that first order
wavefunction has been optimized but it still has a pair product
form. Using Eq. (\ref{usiter}),  neglecting constants and
combining pair terms together, we get in second order a function
of the form: \be U^{(2)} = {\tilde U}^{(1)} - \langle\langle
\sum_i \lambda_i (\nabla_i U^{(1)})^2 \rangle\rangle_1
\label{fku1}
 \ee and \be S^{(2)} = \langle\langle
\sum_i \bfk_i \cdot (\bfr_i - 2 \lambda_i \nabla_i U )
\rangle\rangle_1 . \label{fks1}\ee Here ${\tilde U}^{(1)}$ includes
additional pair terms.

At second order, we cannot perform the averaging analytically,
since it involves drift under the influence of the first order
wavefunction: $U^{(1)}$. We make the assumption that the averaging
will not change the functional form of the quantity being averaged
but only smoothes out the individual functions. That is, our
ansatz for the iterated wavefunction is: \be U^{(2)} = U(R)
-\sum_i (\nabla_i W)^2 \label{fku2}\ee and \be S^{(2)} = \sum_i
\bfk_i \cdot (\bfr_i -\nabla_i Y) \label{fks2}\ee where $~U$, $W$
and $Y$ are three different pair ``potentials'' to be optimized.
In the following, we have adopted the convention that pair
functions have the same sign as $v_{ij} (r)$, so that, for
example, a repulsive $v$ leads to a repulsive $w$ and $y$.

The two new functions appearing at second order are the backflow
function $Y$ and the three-body or polarization term $W$. The
backflow potential is \be Y = \sum_{i<j} y_{ij}(r_{ij} ) \ee where
$y(r) $ is a spherically symmetric function and the sum extends
over all pairs of particles, including both electrons and protons.
The backflow displacement is defined as the gradient of the
backflow potential with respect to a particle coordinate: \be
\Delta \bfr_i =- \nabla_i Y = \sum_{j\ne i} \eta (r_{ij}) (\bfr_i-
\bfr_j ) \label{eta}\ee where \be \eta(r) = -\frac{1}{r}\frac{dy (r)} {dr}\ee
corresponds to the definition in previous work for homogeneous
systems \cite{kwon93,kwon98}.

With this ansatz, the antisymmetized trial function is a
determinant composed of ``quasi-particle'' coordinates: \be
\psi_F^{(2)} = \det \left[ \exp i\bfk_j ( \bfr_i +\Delta \bfr_i )
\right]e^{-U(R) +(\nabla W)^2}. \label{BF3B} \ee Recall that in
the fixed-node or fixed-phase diffusion Monte Carlo method, one
obtains the exact energy subject to the imposed constraint
\cite{RMPQMC,ortiz93}. The assumed node or phase limits the
ultimate accuracy for fermion systems. Since the correction to the
real part, the three-body term, is already symmetric, it is the
backflow which is responsible for the change of node or phase of
the trial function and is, in that sense, more important than the
Jastrow and polarization part.

In the above derivation we have neglected any effects of a complex
drift velocity. However, as already shown in Ortiz and
Ceperley\cite{ortiz95}, a complex drift velocity does not affect
the corrections to the wavefunction to the order we have
considered; the Eqs~(\ref{usiter}) are valid to improve the
wavefunction.

Now we consider the long range properties of the pair functions
appearing. In periodic boundaries (or ``super-cells'') we need to
perform Ewald summations of the functions $V,U,W,Y$. This is most
convenient in Fourier space. We define the Fourier transform of a
radial function as: \be {\tilde y}_k = \int d\bfr^3 e^{-i
\bfk\bfr} y(r). \ee Using the Poisson sum formula, the
``potential'' of the $i^{th}$ particle in periodic boundary
conditions is: \be y_i = \frac{1}{\V}\sum_{\bfk,j} {\tilde y}_k
e^{i\bfk(\bfr_i- \bfr_j) } \ee  where $\V$ is the volume of the
supercell. For example, to find the backflow displacement,
Eq.(\ref{eta}), we simply take the gradient of the pair function:
\be \Delta \bfr_i = - \frac{1}{\V}\sum_{\bfk,j} i\bfk {\tilde y}_k
e^{i\bfk(\bfr_i- \bfr_j) } \ee  where $\bfk$ ranges over the
reciprocal lattice vectors of the supercell.

The three-body potential, $W$ is defined analogously in terms of a
pair polarization $w(r)$.  This function is related to that used
in previous QMC work\cite{panoff89,moroni95,kwon93,kwon98} by: \be
\sqrt{|\lambda_T|} \xi (r) =\frac{1}{r}\frac{dw (r)} {dr}. \ee The
overall sign of $w$ is not important because only its square
appears in the trial function, but the relative sign of the
electron-electron to the electron-proton interaction is
significant.

\begin{figure}
\centerline{\psfig{figure=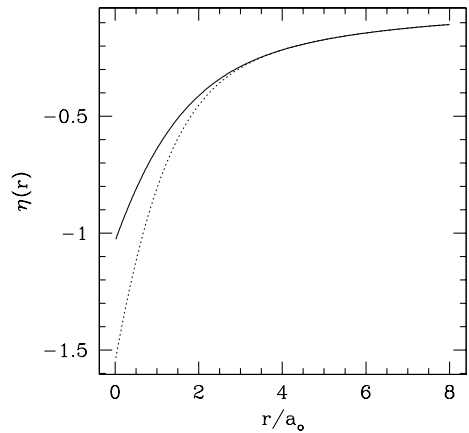,width=8cm,angle=0}} \caption{
The $u^{ep}(r)$ using the RPA (Gaskell) form (dotted line) and
$\eta(r)$ (solid line) from smoothing it with  $k^{-2}$ for the
e-p correlation at $r_s$=1.31, both computed for an infinite
system. Note that in this approximation
they both tend to the same limit at large $r$. }
\label{yofr}
\end{figure}

One of the simple ways of deriving conditions on the backflow
function is to look at the action of the Hamiltonian on the
wavefunction, the local energy, and to minimize the fluctuations
of the local energy. Here we focus on the imaginary part of the
local energy and consider a single electron with phase $S= \bfq
\cdot( \bfr - \nabla Y)$. Setting to zero the imaginary part of
the local energy we obtain: \be \nabla \nabla^2 y(r) +2 \nabla
u(r)-2 \nabla u(r)\nabla \nabla y(r) =0.\ee Neglecting the last
term, since it is higher order in the interaction, we obtain:
$\nabla^2 y(r) = -2 u(r) .$ This has a solution in Fourier space:
$ y_k = 2 u_k/k^2 $. (Because we want a solution which is smoother
than $u(r)$ at $r=0$, we neglect a term proportional to $r^{-1}$.)
We get the same smoothing ($k^{-2}$) that we observed at first
order for the pair function. Shown in figure \ref{yofr} are the
$u(r)$ and $\eta(r)$ function coming from this approach.
Note that this approach is based on a single electron description and
therefore  does not correctly describe the long wavelength
(large r) behavior where the collective motion dominates.

To obtain a simple form for the three-body potential, we note that
the averages used in the definition of $Y$ are similar to those
for $W$, see Eq.(\ref{fku1}-\ref{fks2}).  Hence an estimate of the
polarization potential is  \be W=-\sqrt{\lambda}k_F Y \ee where we
have approximated $\langle\langle ( \nabla_i U )^2 \rangle\rangle
\approx (\langle \nabla_i U \rangle)^2/\tau $, averaged over a
``typical'' time $\tau \approx (\lambda k_F^2)^{-1}$. This relates
the three-body contribution to the backflow potential.

The GFK approach is good for suggesting corrections, but there are
serious problems in using it to find a good backflow function
since the averaging is difficult to carry out, the linear cumulant
approximation may be inadequate, and the long-time effects of the
imaginary drift are being ignored. If one cannot analytically
perform the averaging, one does not know what time to multiply the
local energy by to get a wavefunction, nor the relative
corrections at large versus small distances.  We now discuss
several other approaches which allow us to directly evaluate the
Jastrow, threebody and backflow functions and give more insight
into their form.

\section{Perturbation Theory/Analytic methods}

In this section we follow another approach to obtaining improved
estimates of the many-body wavefunction. Many-body perturbation
theory is a well studied approach to understanding the effects of
weak correlation. Encouraged by the use of the RPA \cite{gaskell}
which gave an excellent analytic two-body correlation function, we
will extend this wavefunction by perturbative expressions for the
Jastrow, backflow and threebody potentials for the electron gas
and for metallic hydrogen. Rather than performing a systematic low
or high density expansion to derive analytical expressions for the
variational wavefunction of the electron gas or metallic hydrogen,
we concentrate on improving this correlation factor. The
collective coordinate formulation of Bohm and Pines
\cite{BP,gaudoin} allows us to use Slater-Jastrow wavefunctions as
zero$^{th}$-order starting point. We obtain improved potentials
for the homogeneous electron gas and metallic hydrogen, which
compare very well with numerically optimized forms.

Even if perturbation theory assumes a weak coupling (or
high density) expansion, we expect the derived properties to be
qualitatively valid as long as the corresponding perturbation
expansion remains regular, {\it e. g.} until there is a phase
transition to an insulating phase.

\subsection{Single particle perturbation theory}

Consider a single electron interacting with an arbitrary external
potential $v(\bfr)$ with Fourier transform $\tilde{v}(\bfk)$. To
avoid the problems arising from the long-range behavior of the
Coulomb interaction, we restrict the analysis to a potential with
a Fourier transform which is finite at the origin, $|\tilde{v}(0)|
< \infty $, {\it e.g.} a screened Coulomb potential. We use the
continuum notation in this section ($\frac{1}{V} \sum_k
\leftrightarrow \int \frac{d^3 k}{(2\pi)^3}$). The solution of the
Schr\"odinger equation $\phi_k({\bf r})$ of a particle with
wavevector $\bfk$, \be \phi_k({\bf r}) = \frac{1}{(2 \pi)^3} \int
d^3 \bfp \, c_k(\bfp) \exp[i \bfp \cdot \bfr] \ee can be written
as
\begin{equation}
c_k({\bf p})= (2 \pi)^3 \delta({\bf k} - {\bf p}) +\frac{4 \pi
f({\bf k},{\bf p})}{k^2 - p^2 +i \delta}
\label{ck}
\end{equation}
where the off-shell scattering amplitudes, $f({\bf k},{\bf p})$,
are given by the integral equation
\begin{equation}
4 \pi \lambda f({\bf k},{\bf p}) = \int \frac{d^3 k'}{(2 \pi)^3}
\tilde{v}({\bf p}-{\bf k}') c_k({\bf k}'). \label{fk}
\end{equation}
Using the Born approximation we can write down the wavefunction to
first order in $\tilde{v}$
\begin{eqnarray}
\phi_k({\bf r}) & \simeq &
\phi_k^{(0)}({\bf r}) + \phi_k^{(1)}({\bf r})
\nonumber \\
& = & e^{i {\bf k} \cdot {\bf r}} \left( 1-\frac{1}{\lambda} \int
\frac{d^3 p}{(2 \pi)^3} e^{i {\bf p} \cdot {\bf r}}
\frac{\tilde{v}(p)}{{\bf p} \cdot ({\bf p} + 2 {\bf k})} \right)
\end{eqnarray}
If we expand the solution around ${\bf k }=0$ and assume the
change in the wavefunction is small, we can write it in the
pair-product and backflow form, Eq.(\ref{BF3B}). We obtain for the
pair potential
\begin{equation}
u({\bf r})\simeq \int \frac{d^3 p}{(2 \pi)^3} e^{i {\bf p} \cdot
{\bf r}} \frac{\tilde{v}(p)}{\lambda p^2},
\label{upert}
\end{equation}
and for the backflow potential
\begin{equation}
y({\bf r})\simeq 2 \int \frac{d^3 p}{(2 \pi)^3} e^{i {\bf p}
\cdot {\bf r}}  \frac{\tilde{v}(p)}{\lambda p^4}.
\label{ypert}
\end{equation}
Note that the small ${\bf p}$ part of the integral is usually cut
off by the finite size of the box. In addition, it is $\eta(r)$,
the derivative of $y(r)$ (see Eq. (18)) which enters in the trial
function.

Although the first order approximation is only reliable in the
case of a weak potential, it becomes correct in the high momentum
region and hence, gives the correct cusp conditions. The derived
form is identical to that obtained from the Feynman-Kacs formula
in the previous section. For an arbitrary weak potential, we
further get the long range behavior, $u \propto \tilde{v} (0)/r$
and $\eta \propto 1/r^3$ for $r \to \infty$, provided the
potential has a finite range ($\tilde{v}(q) - \tilde{v}(0) \propto
q^2$ for $ q \to 0$) and there is no other singularity in the
integrand.

To find an approximate form for the three-body function $W(\bfr)$
we must go to higher order in the interaction, but only at $\bfk
=0$. Using Eqs.~(\ref{ck}-\ref{fk}), we can write down the second
order corrections in $\tilde{v}$ to the wavefunction,
$\phi_k^{(2)}$ at $k=0$:
\begin{equation}
\phi_{k=0}^{(2)}({\bf r})=
\frac{1}{\lambda^2} \int \frac{d^3q}{(2\pi)^3}
\frac{d^3p}{(2\pi)^3} \frac{\tilde{v}(q) e^{i{\bf q} \cdot {\bf
r}}}{({\bf q}+ {\bf p})^2 } \frac{\tilde{v}(p) e^{i{\bf p} \cdot
{\bf r}}}{p^2 } \nonumber
\end{equation}
\begin{equation}
\approx  \int \frac{d^3q}{(2\pi)^3} \frac{d^3p}{(2\pi)^3}
\frac{\tilde{v}(q) e^{i{\bf q} \cdot {\bf r}}}{\lambda q^2 }
\left( 1- \frac{ 2 {\bf q}\cdot {\bf p}}{q^2} \right)
\frac{\tilde{v}(p) e^{i{\bf p} \cdot {\bf r}}}{\lambda p^2 }.
\end{equation}
This is almost in the form of the three-body correlation obtained
with the FK approach: $(\nabla w)^2$. Note, however, that Eq. (33)
is unsymmetrical in $q$ and $p$ so that in r-space it will be
written as: $(\nabla w_u) \cdot (\nabla w_y)$ with $w_u({\bf r})
\simeq u(r)$ and $w_y({\bf r}) \simeq y({\bf r})$. Therefore the
polarization term is not a square but a product of the gradients
of two different functions. (In second order one will also find a
contribution $\propto [u(r)]^2$ to the pair term.)

The perturbative expressions (\ref{upert}) and (\ref{ypert}) are
based on the Born approximation for scattering between free
states. However, an attractive potential as the electron-proton
(effective) interaction might also lead to bound states. To
include the effects of a possible bound state we can use the
non-perturbative expression (\ref{fk}) for the scattering
amplitudes: given an approximate expression for the bound state
wavefunction of energy $\epsilon_k=-\lambda k^2$, we can calculate
the scattering amplitudes and obtain corrections from the bound
state to the pair and backflow potential in the same way as shown
above for the scattering states within the Born approximation. In
a similar way one should proceed to obtain approximations for the
pair and backflow potentials for systems where the interatomic
potentials cannot be treated within the Born-approximation, for
example potentials dominated by a hard core.

Of course, in the case of a single electron in an external
potential we can solve the Schr\"odinger equation by other means
and obtain the ``best'' pair and backflow potentials from the
exact (numerical) solution. This is done below for a perfect
crystal using a band structure calculation. However, the simple
perturbative approach above provides an easy way to get some
intuition for the pair and backflow potential, and is already good
enough to determine their asymptotic properties. These properties
are expected to hold in the many-body case: the short-range
properties are typically determined by two-body collisions and the
influence of the remaining particles on the long-range properties
are usually well described by an effective single particle
potential. Many-body perturbation theory, which we discuss next,
leads to similar expressions.

\subsection{Many-body perturbation theory}

We now make an expansion of the exact $N$ particle wavefunction
$|\phi \rangle$ of the interacting system around the
non-interacting (ground) state $|\phi_0 \rangle$; the ground state
without both the electron-electron and electron-proton
interaction. Let $a_k$ ($a_k^\dagger$) be the annihilation
(creation) operator for an electron of wavevector $k$. Expanding
in particle-hole excitations, we have:
\begin{equation}
|\phi \rangle  \propto \big(  1  +
 \sum_{q,k_1,k_2} \alpha_{k_1,k_2,q}
a^\dagger_{k_2-q} a_{k_2} a^\dagger_{k_1+q}a_{k_1} + \dots \big) |
\phi_0 \rangle.  \label{p1}
\end{equation}
The problem is reduced to determining the coefficients
$\alpha_{k_1,k_2,q}$. Just as in the single particle case, a
further expansions of $\alpha_{k_1,k_2,q}$ around $k_1=0$ or
$k_2=0$ together with an exponentiation brings the wavefunction
into the desired functional form, thereby determining the pair and
backflow potentials. To avoid over counting, we assume that the
summation in Eq.(\ref{p1}) goes only over distinct states so that
it is sufficient to antisymmetrize the wavefunction at the very
end, once we have calculated the perturbative corrections. We have
limited the expansion in Eq.(\ref{p1}) to the leading order
corrections, particle-hole excitations; the generalization to
include higher order excitations is straightforward, but not
necessary to calculate the pair and backflow terms in the
wavefunction.

In order to determine the coefficients, we write $|\Phi_{k1,k2,q}
\rangle =a^\dagger_{k_2-q} a_{k_2} a_{k_1+q}^\dagger a_{k_1}
|\phi_0 \rangle$ and multiply these states by a constant phase
\begin{equation}
\alpha_{k1,k2,q} = \langle \Phi_{k1,k2,q}   |\phi \rangle
\propto\frac{ \langle \phi |  \phi_0 \rangle \langle
\Phi_{k_1,k_2,q}^{ee} | \phi \rangle}{ \langle \phi | \phi_0
\rangle\langle \phi_0 | \phi \rangle} \label{alphas}
\end{equation}
Note that with this phase factor, the right hand side of
Eq.~(\ref{alphas}) is given by the expectation value of an
operator over the true ground state and the coefficients can
therefore be identified as a $N$-particle Green's function
\cite{Greens}. By considering only particle-hole excitations in
Eq.(\ref{p1}), the $N$-particle Green's function reduces to a
connected two-particle Green's function and the lowest order
modifications to the ideal gas ground state of the homogeneous
electron gas are therefore related to the two particle Green's
function $G^{(2)}$ at equal times \cite{Greens} or equivalently
\begin{equation}
\alpha_{k1,k2,q} \simeq
\frac{ \langle \phi |
a^\dagger_{k_2-q} a_{k_2}
a_{k_1+q}^\dagger a_{k_1} |\phi \rangle }{\langle \phi | \phi \rangle}
\end{equation}
Summing up particle-hole bubble diagrams (corresponding to the RPA
approximation) results in an effective interaction,
$\tilde{v}_{RPA}(p,\omega)$, \be \tilde{v}_{RPA}(p,\omega)=
\frac{\tilde{v}(p)}{\epsilon(k,\omega)}, \quad \epsilon(k,\omega)=
1-\tilde{v}(p)D(p,\omega) \label{vrpa} \ee where $D(p,\omega)$ is
the Lindhard function. Perturbation theory can now be arranged to
be regular \cite{Gell-Mann}. We note that Eq. (\ref{vrpa}) already
contains the correct short- and long-range limit of the effective
interaction.

Neglecting for the moment any contributions from plasmon excitations
coming
from the poles where $\epsilon(k_p,\omega_p(k_p))=0$,
we get
\begin{eqnarray}
\alpha_{k_1,k_2,q} & = &
(1-n_{k_1-q})n_{k_1}(1-n_{k_2+q})n_{k_2}
 \label{pert}\\
&\times &
\frac{\tilde{v}_{RPA}(q,\varepsilon_{k_1}-\varepsilon_{k_1-q})
+\tilde{v}_{RPA}(q,\varepsilon_{k_2}-\varepsilon_{k_2+q})}
{2(\varepsilon_{k_1}+\varepsilon_{k_2}-\varepsilon_{k_1-q}-\varepsilon_{k_2+q})}
\nonumber
\end{eqnarray}
where $n_k$ are the occupation numbers of state $k$ in lowest order. Expanding
around $k_1=k_2=0$, we get the Jastrow and the backflow potential.
Including the plasmon excitations will give an important
long-range contribution. However, in the simplest approximation,
this contribution describes only the long wavelength limit
correctly, and destroys the correct short distance behavior. We
will circumvent this problem in the next section using the
formalism of collective coordinates.

As already shown in the previous section, we expect a more general
form for the three-body potential, \be \phi \propto \det \left[
\exp i\bfk_j ( \bfr_i +\Delta \bfr_i ) \right]e^{-U(R) + W}
\label{BF3Busym} \ee with W= \be \sum_{j} (\nabla_j W_u) (\nabla_j
W_y) - 2 \sum_{i<j} [\nabla_j w_u(r_{ij})] [\nabla_j w_y(r_{ij})],
\label{w1} \ee where \be w_u(r) \simeq u(r) \quad w_y(r) \simeq
y(r). \label{w2} \ee

For the interactions of the electrons with static protons, we can
use the static dielectric function $\epsilon(k,0)$ to obtain the
effective electron-proton interaction, and use directly the
results of the single particle perturbation theory of the previous
section with this screened potential.

The disadvantage of perturbation theory is that one gets correct
behavior at long and short distances, but it does not provide an
unique way to interpolate between these limits. In Table
\ref{asym} we summarize the asymptotic properties of the pair and
backflow potentials for the 3D electron gas.

\begin{table}
\begin{tabular}{|c||c|c|c|c|}\hline

  function& $r\rightarrow 0$ & $r\rightarrow \infty$ &
$k\rightarrow 0 $ & $k\rightarrow \infty$\\\hline\hline
 $v$ & $e^2/r$ & $e^2/r$ & $4 \pi e^2 / k^2$ & $4 \pi e^2 / k^2$
 \\\hline

 $u$ & $u_0-\frac{e^2 r}{4\lambda}$ & $\sqrt{\frac{e^2}{8\pi n \lambda}} \frac{1}{r}$ & $\sqrt{\frac{v_k}{2n\lambda
 k^2}}$ & $\frac{v_k}{2\lambda k^2}$\\\hline

 $y$ & $y_0-y_2r^2+\frac{e^2}{48 \lambda}r^3$ & $\frac{c(r_s)}{4 \pi n r}$ &
 $\frac{c(r_s)}{n k^2}$& $\frac{v_k}{2\lambda k^4}$ \\\hline
  $\eta$ & $2 y_2- \frac{e^2}{16 \lambda} r$  &  $\frac{c(r_s)}{4 \pi n r^3}$ & & \\
 \hline
\end{tabular}
\caption{ Asymptotic properties of the Jastrow and backflow
functions for the 3D electron gas. $\lambda=\hbar^2/2m$, $n$ is
the electron density, $y_2 \approx 0.055 r_s$, and $c(r_s)\approx
1+ 0.075 \sqrt{r_s}/(1+0.8 \sqrt{r_s})$. }\label{asym}
\end{table}

\subsection{The Bohm-Pines collective coordinate approach}

Instead of replacing the established form for the Jastrow part
proposed by Gaskell \cite{gaskell} by the direct use of Eq.
(\ref{pert}), we prefer to improve the RPA form of Gaskell
by extending it using
perturbative formulas. This is most easily done within the
framework of the collective coordinate description of Bohm and
Pines using additional field variables \cite{BP}. In this
approach, the original Hamiltonian of electrons interacting with
each other and with static protons is extended by an additional
boson field with generalized momentum variables $\Pi_k$ coupling to
the electron and proton density fluctuations
\begin{eqnarray}
H& =&\sum_i \lambda p_i^2+\frac{1}{2\V} \sum_k \tilde{v}_k \left(
\rho_{-k}^{e} \rho_k^e-N \right)
\nonumber \\
& & - \frac{1}{\V} \sum_k \tilde{v}_k \, \rho_{-k}^{e}
\rho_k^p
\nonumber \\
& &+\frac{1}{\V}\sum_k \left( \frac{\Pi_k^\dagger \Pi_k}{2} + M_k
\Pi_k^\dagger \rho_k^e + P_k \Pi_k^{\dagger} \rho_k^p \right)
\end{eqnarray}
where $\rho_k^e$ ($\rho_k^p$) is the Fourier transform of the
electron (proton) density, $\rho_k=\sum_i e^{-i{\bf k}\cdot {\bf
r}_i}$, and $M_k$ and $P_k$ are variational parameters. By
imposing the extra conditions $\Pi_k \Psi =0$ on the wavefunction,
the ground state wavefunction of the new extended Hamiltonian will be
identical to the original one. For a detailed description of this
approach we refer to the original literature \cite{BP}; we will
only describe the main steps.

Carrying out the following canonical transformation \be
\phi_{old}=\exp[iS/\hbar] \phi_{new}, \quad S=\frac{1}{V}\sum_k
\left( M_k \rho_k^e +P_k \rho_k^p\right) Q_k, \ee where $Q_k$
represents the field coordinate conjugate to $\Pi_k$, we obtain an
equivalent Hamiltonian
\begin{eqnarray}
H& = &\sum_i\lambda_e p_i^2 + \frac{1}{\V} \sum_k \left(
\frac{\Pi_k^\dagger\Pi_k}{2} +\lambda_e n k^2 M_k^2 Q_k^\dagger
Q_k \right)
\nonumber \\
& & +H_{sr}^{ee} + H_{int}  + H_{rw} +H_{sr}^{ep} \label{BP11}
\end{eqnarray}
where
\begin{eqnarray}
H_{sr}^{ee}& = &\frac{1}{2\V}\sum_k \left(\tilde{v}_k-M_k^2
\right) \left(\rho_{-k}^{e} \rho_k^e -N \right) \label{eeint}
\\
H_{int} & = & i \frac{1}{\V} \sum_{k,j} \left( \frac{ {\bf k}
\cdot {\bf p}_j }{m} + \frac{ \hbar k^2}{ 2m} \right) M_k Q_k
e^{-i {\bf k} \cdot {\bf r}_j} \label{eplasi}
\\
H_{rw} & =& \frac{\lambda}{\V^2} \sum_{k \ne k',j} Q_k^\dagger
Q_{k'} M_k M_{k'}  {\bf k} \cdot {\bf k}'  e^{i({\bf k}-{\bf
k}')\cdot {\bf r}_j} \label{rotat}
\\
H_{sr}^{ep}& = &-\frac{1}{\V}\sum_k \left(\tilde{v}_k-M_kP_k
\right) \rho_{-k}^e \rho_k^p \label{epint}
\end{eqnarray}
Now the ground state of the additional field in the zero$^{th}$
order Hamiltonian, Eq.(\ref{BP11}), is simply given by harmonic
oscillator ground states of frequencies $\V_k=(n k^2
M_k^2/m)^{1/2}$, \be \phi_{new}^{0}= det[ \exp[ i {\bf k}_i \cdot
{\bf r}_j ] ] \exp\left[ -\frac{1}{\V}\sum_k \frac{ \Pi_k^\dagger
\Pi_k}{2 \hbar \V_k} \right] \ee Transforming back and applying
the subsidiary conditions replaces the field operator $\Pi_k$ by
$M_k \rho_k^e+P_k \rho_k^p$ and the zero$^{th}$ order wavefunction
is in the Slater-Jastrow form
\begin{eqnarray}
\phi_{old}^{0}& = & det [ \exp[ i {\bf k}_i \cdot {\bf r}_j ]]
\nonumber \\
&& \exp\left[ -\frac{1}{\V}\sum_k \frac{M_k^2 \rho_{-k}^e
\rho_k^e +2 M_k P_k \rho_{-k}^e \rho_k^p }{2 \hbar \V_k} \right]
\end{eqnarray}
up to a constant factor. Instead of using $M_k=(\tilde{v}_k)^{1/2}
\theta(k_c-k)$ for the long wavelength part up to $k_c$, and
optimizing the cut-off $k_c$, as done in the original work of Bohm
and Pines, we can use $\tilde{u}_k^{ee}$ and $\tilde{u}_k^{ep}$
for the electron-electron  and electron-proton Jastrow part taken
in the RPA approximation\cite{dmc81} and relate these functions to
$M_k$ and $P_k$. The resulting residual electron-electron and
electron-proton interaction is screened, since $M_k^2 \to
\tilde{v}_k$ and $M_k P_k \to \tilde{v}_k$  in the long wavelength
limit $k \to 0$.

A second unitary transformation using \be S=\frac{1}{\V}
\sum_{k,j} M_k \frac{{\bf k}\cdot {\bf p}_j}{m \omega_p(k)( \hbar
\omega_p(0)+\epsilon_k)} \Pi^\dagger_k e^{-i{\bf k} \cdot{\bf
r}_j} \label{S1} \ee eliminates $H_{int}$ to first order. Here,
$\omega_p(k)$ is the plasmon frequency at wavevector $k$. Note
that this transformation brings the wavefunction into the backflow
form. Furthermore, we treat the remaining terms of the Hamiltonian
perturbatively as shown in the previous subsection.

The detailed functions we used for the electron gas and for
metallic hydrogen are given in the appendix and the numerical
tests are given in Section V.

\section{Comparison with the Band Structure Wavefunction}

In this section, we consider another approach of generating
backflow functions.  As in the discussion of the single particle
perturbation theory in the last section, we consider a perfect
lattice of protons in which a single electron moves. It is
straightforward to expand the wavefunction in plane waves and
obtain a precise numerical solution  of the one electron problem
by diagonalization of the Hamiltonian matrix. We study to what
extent we can recast the ``band structure'' wavefunction into a
backflow form. The advantage of this approach is that we are
evaluating the entire non-linear effect of a lattice of protons on
the electron wavefunction, or orbital, which for a perfect lattice
is a Bloch wave. However, effects of electron correlation or
screening are absent for this model.

As was done in Eq. (26), the exact one-electron wave function is
expanded in plane waves: \be \phi_{\bfk}(\bfr) = \sum_G c_{k,G}
e^{i(\bfG+\bfk)\cdot\bfr}\ee where $\bfG$ is a reciprocal vector
of the lattice and $\bfk$ the crystal momentum. We then obtain
numerical values for $c_{k,G}$ by conventional diagonalization of
the Hamiltonian in this basis.

First, we study the wavefunction at $\bfk=0$ to determine the pair
part of the wavefunction, $U$. Neglecting the three-body term we
have: \be \sum_i u(|\bfr - \bfz_i| )= -\ln(\phi_0 (\bfr)) \ee
where $\bfz_i$ are the proton positions. Then by fourier
transforming and assuming a Bravais lattice : \be u_{\bfG} = -
\int_{\V} d^3 r e^{-i \bfG\bfr} \ln(\phi_0 (\bfr)).\ee This is
shown in Fig. \ref{jastrow} and compared to the RPA form (solid
line) and cumulant form. Note that we only obtain information
about $u_{\bfk}$ at values of $\bfk$ on the reciprocal lattice. It
is seen that except for the first few reciprocal lattice vectors,
the pair wavefunction is determined by the cusp behavior. The
non-cusp behavior is due to the neglect of higher order terms in
the cumulant expansion. Some effects are picked up by the
three-body term of the wavefunction. We note that even for the
largest lattice vector, the values seem to follow a smooth curve,
independent of the lattice directions. The $\bfk=0$ component,
though important, will not affect the many-body nodal structure or
the correlation effects near the fermi surface.

Now, let us use the same procedure to estimate the backflow
function. First, we divide out the wavefunction at $\bfk=0$ to
define the backflow functions : \be \bfk \cdot \nabla Y_k(\bfr)
=\bfk \cdot \bfr + i \ln[\phi_{\bfk}/\phi_o].\ee  Assuming
$Y_k(\bfr)$ is the sum of contributions of proton-electron terms
on a Bravais lattice we get: \be \bfk \cdot \nabla Y_k(\bfr)
=\frac{1}{\V}\sum_{\bfG} i\bfG \cdot \bfk y_{\bfG}^k e^{i
\bfG\cdot\bfr}.\ee Setting these two expressions equal and taking
the Fourier transform we arrive at: \be y_q^k =
\frac{-i}{\bfk\cdot\bfq} \int d\bfr e^{-i\bfq\bfr} \ln[\phi_{\bfk}
(\bfr) / \phi_o (\bfr)]. \ee

In general, the function $y_q^k$ depends on both $\bfk$ and
$\bfq$. For small values of $k$, the ratio approaches a limit,
independent of both the magnitude and direction of $\bfk$. As with
the pair term, we can only determine $y_{\bfk}$ at reciprocal
lattice vectors, $\bfq$ and for $\bfk$ in the first Brillouin
zone. Shown in Fig. (3) is the ratio for several values of $\bfk$
evaluated for a simple cubic lattice plotted versus $\bfq$. The
dispersion of the values from a smooth curve is a test of the
extent to which the band structure orbital can be cast into the
form of a backflow function. Note that only for the smallest
values of $\bfq$ is the backflow function appreciable. At
intermediate values of $q$ one does seem some effect of
''non-backflow'' behavior, however it is not clear how important
these effects are. At large $q$, we see the behavior $y_q \simeq 8
\pi e^2 /\lambda q^6$ shown as the solid line as expected from the
results of Section II and III.

\begin{figure}
\centerline{\psfig{figure=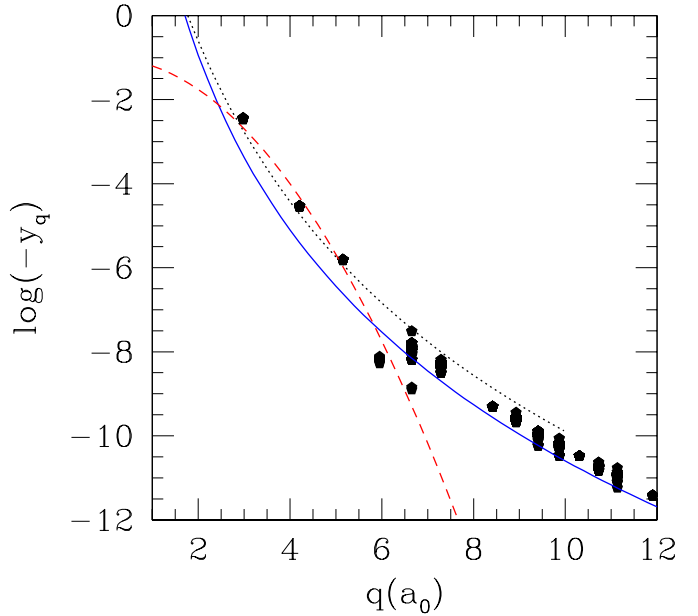,width=10cm,angle=0}}
 \caption{
The backflow function $y_q$ versus the wavevector $q$~ in atomic
units for solid cubic hydrogen lattice at $r_s=1.31$.  The solid
symbols are computed using different values of $\bfk$ and $\bfq$in
the range of 0.01 to 0.1 using band theory and Eq. (56). The solid
line is the cumulant approximation: $y_q = -16\pi/q^6$. The dashed
line is the backflow function optimized for an interacting $N$
body hydrogen with a Gaussian form. Dotted line is from Eq. (69).
\label{logy}}
\end{figure}

\begin{figure}
\centerline{\psfig{figure=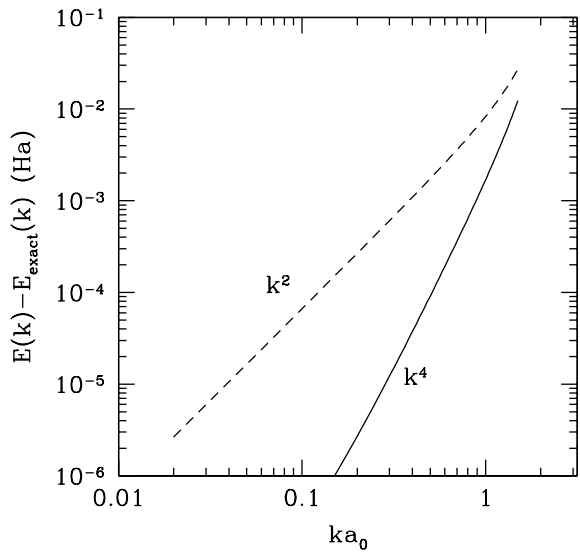,width=10cm,angle=0}} \caption{
The error in the band energy of a single electron in a bcc proton
lattice for $r_s=1.31$ as a function of $k$ (in the 100 direction)
using plane waves (dashed line) and using a BF function (solid
line). Both approximations are exact at the $\Gamma$ point since a
trial function exact at $\bfk$ was used, but for the BF trial
function, the error $\propto k^4$ while in the PW case (zero
backflow) the error is $\propto k^2$\label{bande} }
\end{figure}

In figure \ref{bande} is shown the error in the band energies with
a backflow wavefunction (BF) and the results for having no
backflow effects. For the comparison we used a BF function $y_q =
y_0 \exp(-b q) $ fitted to the low $q$ behavior. Since $y_q$ drops
off rapidly w.r.t $q$, it is primarily the effects at small $k$
that are important to describe\cite{footnoteBT}. By definition the
energies are identical at $k=0$ and the curvature around $\bfk=0$
is exactly put in by the backflow ansatz, at least assuming cubic
symmetry. We see that the errors in the band energy go as $k^4$
instead of $k^2$ for the non-backflow trial function.  However,
near the band edge there are serious problems because our assumed
form does not have mixing of the bands required by lattice
periodicity. We expect such an effect to be much reduced for a
disordered system since such degeneracies will not occur.

This achieves our goal or showing that the dominant band structure
effects can be interpreted as backflow corrections, particularly
at small $\bfk$. This implies that the changes in the nodal
surfaces due to an external potential of protons are well
approximated by backflow functions.

The backflow form is a much more succinct description of the
single body wavefunction than the expansion in plane waves. In the
introduction, we emphasized that this improves performance because
we no longer have to perform the band structure calculation.
However, there is also an improvement in speed of calculation of
the orbitals using backflow.  The expansion in plane waves can be
quite slow, since the accuracy versus number of terms decreases
quite slowly. In previous work on metallic hydrogen
\cite{natoli93}, we divided the band structure orbital by an
electron-proton Jastrow factor as an approximation to $\phi_0
(r)$, and then re-expanded in plane waves. The resulting expansion
is much more quickly convergent in the number of plane waves since
the cusp at $r_{ep}=0$ is in the Jastrow factor. It takes the sum
of many plane waves to recover this non-analytic behavior at
$\bfr_i=\bfz_j$. Backflow takes this even further by using the
fact that near $\bfk = 0$ the wavefunction can be expanded in pair
terms with a higher-order cusp. These pair terms can be
conveniently  and rapidly computed, since much of the
computational effort is to map each pair of particles (ee or ep)
onto a grid value for a table look-up. The distances and grid
values are then used for all of the pair terms: the potential, the
Jastrow, the backflow and the polarization terms.

The problems concerning degeneracies of the unperturbed plane wave
functions near the edge of the Brillouin zone are common to all
analytical approaches considered up to now. Without a separate
treatment of (nearly) degenerate zeroth order (plane wave) states,
neither the cumulant method (Section II) nor perturbation theory
(Section III) are able to produce the resulting energy splitting
at the band edge. A degenerate case will have to be treated by
including all of the degenerate states in the unperturbed basis.

\section{Quantum Monte Carlo Testing of Trial Function Forms}

There are two principal simulation methods used to calculate the
ground state energies of quantum many-body systems: Variational
Monte Carlo (VMC) and Diffusion Monte Carlo (DMC). In VMC, one
samples the square of the wavefunction, and, in DMC, one uses a
trial wavefunction and the imaginary-time evolution to project
onto the ground state. VMC is potentially very powerful because
one can use any wavefunction, as long as one can easily compute
its values. One can add  correlation directly to the wavefunction,
leading to a very compact accurate wavefunction. The resulting
integrals are similar to that of the classical partition function
and therefore demand a simulation algorithm to evaluate. The
disadvantage of the variational approach is that one needs to use
the right functional space in order to get satisfactory
properties. Though DMC is much less dependent on details of the
trial wavefunction than VMC, however, lacking an exact fermion
algorithm, the results still depend to some extent on the
positions of the node (or phase) of the trial wavefunction.

The most straightforward, and rigorous approach to determine the
trial function is to propose a definite analytic form, containing
some parameters, $a$. One then uses VMC to evaluate the
variational energy $E_V (a)$, an upper bound to the exact energy
as a function of $a$. One can use various techniques to optimize
the parameters to obtain the lowest energy, the lowest variance or
some combination of the two. Variational
optimization\cite{kwon93,kwon98} has determined good backflow and
three-body trial functions for the electron gas in both 2 and 3 D
. The disadvantage of optimization is that beyond general trends,
it is hard to extract analytic behavior because of the noisy
behavior of the optimization method and the restriction to a
limited functional form.

Here we compare several different trial wavefunctions on two
systems: the 3D electron gas and metallic hydrogen. We employ
three estimators of the quality of the wavefunction: the
variational energy $E_v = \langle \psi \calH \psi \rangle$, the
variational variance $\sigma^2=\langle \psi \calH^2 \psi
\rangle-E_v^2$ and the DMC (fixed-node) energy. The first two
properties are sensitive to all aspects of a wavefunction; the
variance is particularly sensitive to short-range structure since
the energy fluctuations are larger. However, the DMC energy is
determined only by the positions of the trial function node, not
by the ``bosonic'' part of the trial function.  The VMC/DMC
calculations  we performed were standard ones\cite{RMPQMC}. All
calculations are done with Periodic Boundary Conditions (PBC),
equivalent to the $\Gamma$ point for a band structure calculation
in a cubic unit cell. Hence all trial functions were real. Though
twist-averaged boundary conditions (TABC) \cite{lin2002} are
useful in reducing size effects, tests showed the relative
accuracy of various trial functions can be determined with PBC
using real trial functions.

First, we discuss the results using backflow and three-body
wavefunctions on the 3D electron gas as shown in Table II. The
results using analytic trial functions give results comparable to
the numerically optimized backflow results of Kwon et
al\cite{kwon98}.  We find that for $r_s <20$ the analytic
wavefunction have a lower VMC energy than the numerically
optimized wavefunction. This is mainly due to the inclusion of the
long range part of the backflow potential. For all values of $r_s$
the analytic wavefunctions have a lower DMC energy, implying a
more accurate nodal surface than obtained by numerical
optimization. For $r_s=20$ the numerically optimized VMC energy is
lower than that of the analytic wavefunction, indicating that at least
the 3-body part of the wavefunction becomes
inaccurate at strong correlations.

\begin{table}
\begin{tabular}{|c|l|l|l|l|} \hline

 $r_s$   & wavefunction & $E_v$& $\sigma^2$ &$E_{DMC}$ \\\hline
  1 &   SJ    &   1.0669 (6) & 1.15 (2)  & 1.0619 (4)   \\
     &   BF3-O    &  1.0613 (4) & 0.028 (1)  & 1.0601 (2)   \\
     &   BF-A   &  1.0611 (2) &  0.029 (1)& 1.0597 (1)   \\
     &   BF3-A   &   1.0603 (2) &  0.022 (1)&    \\\hline
   5  &   SJ    &  -0.15558 (7) &0.0023(1)  & -0.15734 (3) \\
       &   BF3-O    &   -0.15735 (5) & 0.00057 (1)&-0.15798 (4)  \\
       &   BF-A    &   -0.15762 (1) & 0.00061 (1)& -0.15810 (1) \\
       &   BF3-A    &   -0.15773 (1) & 0.00050 (1)& \\\hline
   10  &   SJ    &   -0.10745 (2) &0.00039 (.5)& -0.10849 (2)   \\
        &   BF3-O    &  -0.10835 (2) & 0.00014 (.5)& -0.10882(2) \\
        &   BF-A    &   -0.10843 (2) &0.00017 (1)&  -0.10888 (1)  \\
        &   BF3-A    &   -0.10846 (2) &0.00016 (1)&  \\\hline
   20  &   SJ    &   -0.06333 (1) & 0.000064 (1)  & -0.06388 (1)   \\
        &   BF3-O    &   -0.06378 (2) &0.000027 (7) & -0.06403 (1)\\
       &   BF-A    &  -0.06372  (2) &0.000045 (2) & -0.06408 (1)  \\
       &   BF3-A    & -0.06358   (1) &0.000056 (1) &  \\\hline
\end{tabular}
\caption{ Energies and variances for the 3D electron gas with
$N=54$ unpolarized electrons in Rydbergs/electron. SJ means a
Slater determinant of plane waves times an optimized Jastrow
factor. BF3-O are the result of the numerical backflow-3body
optimization\protect\cite{kwon98}. BF-A are the results using the
RPA Jastrow, Eq.(\ref{uee}) together with the analytical backflow
formula, Eq.(\ref{fullbee}), BF3-A with the additional asymmetric
3-body wavefunction of Eq.(\ref{w1}-\ref{w2}). }
\end{table}

Now we consider the use of these same trial functions for a system
composed of electrons and protons. To determine the properties
using the optimization method, we used the RPA form for both the
electron-electron (ee) and electron-proton (ep) $u(r)$. We used
optimized Gaussians for both the backflow and polarization terms:
\begin{equation} \eta(r)=\lambda
\exp(-(r-r_0)^2/w^2).\end{equation}  Even though the optimal
functions may have a long range tail, as shown earlier, the
additional energy gained is small and we neglect the long-range
terms in setting up the parameterized trial functions.  An
additional Gaussian (with $r_0=0$ so as not to change the cusp
value) was added to the pair term. We did not include ee backflow
or polarization terms in the wavefunction. The resulting
10-parameter wavefunction was then optimized to minimize a linear
combination of its energy and variance. Shown in Fig.~\ref{logy}
are the optimized backflow functions compared with the cumulant
value, with the analytic form and with the band structure
determination. The magnitude and shape are similar, though
differences are apparent.

We compare the results with three other wavefunctions.  The
simplest is the SJ-PW functions\cite{dmc81}, which do not contain
backflow, three-body terms and the orbitals are simple plane
waves. We also used optimized Slater-Jastrow functions with
orbitals from a LDA calculation\cite{natoli93}. Finally shown are
various analytic backflow calculations: one contains only ep backflow
(and 3body),
the others have alse ee backflow (3body) included.

\begin{table}
\begin{tabular}{|c|c|l|l|l|} \hline
 N   & wavefunction & $E_v$& $\sigma^2$ &$E_{DMC}$ \\\hline
  16 &   SJ-PW    &   -0.4754 (2) & 0.0773(25)& -0.4857 (1)   \\
     &   LDA      &   -0.4870 (10)&           & -0.4890 (5)   \\
     &   BF3-O ep   &    -0.4857(1) & 0.0317 (5)& -0.4900 (1)   \\
     &    BF3-A ep&  -0.4798 (1)  & 0.0513 (2)&   \\
     &   BF-A ee+ep&  -0.4850 (1) & 0.0232 (1)&  -0.4905 (1)   \\
     &   BF3-A ee+ep& -0.4850 (1) & 0.0227 (1)&  \\
     &   BF-A ee+ep+b& -0.4878 (1)& 0.0181 (4) & \\\hline
   54  &   SJ-PW    &   -0.5241(3) & 0.0642 (9)  & -0.5329(1)  \\
       &   LDA    &    -0.5365 (5) &             & -0.5390 (5)   \\
       &   BF3-O ep   &  -0.5331 (6) & 0.033 (1)   & -0.5381 (1) \\
       &   BF3-A ep &   -0.5261(1) & 0.0516 (3)   &  \\
       &   BF-A ee+ep& -0.5323 (1) & 0.0222 (2)  & -0.5382(1)  \\
       &   BF3-A ee+ep& -0.5325(1) & 0.0214 (1)  &   \\
     &   BF-A ee+ep+b& -0.5353 (2) & 0.0178 (2) & \\\hline
   128 &   SJ-PW   &   -0.4818 (2) &0.0656 (23)   & -0.4900 (2)   \\
       &   LDA    &   -0.4962  (2) &             & -0.4978 (2)   \\
       &   BF3-O ep  &   -0.4934 (2) & 0.035 (2)   & -0.4958 (3) \\
       &   BF3-A ep &  -0.4846 (3) & 0.059 (1)   &  \\
       &   BF-A ee+ep & -0.4928(2) & 0.030 (1)   &  -0.4978 (4)  \\
       &   BF3-A ee+ep &-0.4926(2) & 0.029 (1)   &   \\
     &   BF-A ee+ep+b&  -0.4947(2) & 0.023(1) & \\\hline
\end{tabular}
\caption{ Energies for bcc hydrogen at $r_s$=1.31. SJ-PW means a
Slater determinant of plane waves times an optimized Jastrow
factor. LDA means LDA orbitals times an optimized 1 body factor
and Jastrow factor\protect\cite{natoli93}, BF3-O ep means optimized
e-p backflow, e-p polarization and Jastrow.  Energies are given in
hartrees per atom. Periodic boundary conditions  ($\Gamma$ point)
and Ewald sums were used. $\sigma$ is the variance per electron.
BF3-A ep are the analytical wavefunctions using ep backflow-3body
only, wheras BF-A ee+ep are results with ee  and ep backflow;
BF3-A ee+ep include also ee and ep 3body and backflow,  BF-A ee+ep+b
uses the same wavefunctions of BF-A ee+ep but
the electron-proton Jastrow and backflow is improved by taking
into account the effects of a bound state.}
\end{table}

Shown in Table III are both VMC and DMC calculations of various
wavefunctions for metallic bcc hydrogen at $r_s=1.31$, a density
very close to the molecular-metallic transition. While the
detailed results depend on the number of particles, in general we
find that the SJ-PW function is in error within VMC by about
15mH/atom while the BF is in error by about 4 mH/atom and the LDA
trial function by about 2 mH/atom. Within DMC the SJ-PW is in
error by 6mH/atom and the BF is as accurate at the LDA
trialfunction within the statistical error. This analysis of
errors is done with the assumption that the LDA-DMC energy is
exact. As another indication of the quality, the VMC wavefunction
variance is roughly a factor of 3 smaller with the BF wavefunction
than with the SJ-PW wavefunction.

We see that for $N=16$ the DMC backflow results are even lower
than the LDA function.  One reason for this could be that $N=16$
has a degenerate ground state for a single Slater determinant;
many-body effects break the degeneracy.  It may be that the
current simulations, though similar to those of Natoli, broke the
degeneracy in a more favorable way and thus have a lower energy.
The $N=54$ system has a non-degenerate ground state at the mean
field level, a closed shell, so the results may be more typical.
Finally degeneracy effects are probably less important at $N=128$
since $N$ is larger.

We have tested the relative importance of including ee backflow in
the case of metallic hydrogen. Using ep backflow-3body only, the
analytical wavefunctions give considerably higher energies
compared to the numerically optimized ones. Including ee backflow
in the analytical forms, they become comparable. One should note
that the analytical approaches derive ee and ep backflow at the
same order of approximation; dropping one of them alone is not
justified and might explain the importance of including ee and ep
backflow in the analytical functions. The inclusion of 3-body
terms does not noticeably affect the energies. This is similar to
the results for the electron gas at comparable
densities\cite{kwon98}. Since the density is close to the
transition from metallic to molecular hydrogen we tried to improve
our wavefunction by considering the effects of a simple
electron-proton bound state on Jastrow and backflow in our
analytical formulas (see appendix) and found significantly lower
energies within VMC.

Note that when ee backflow is included, it becomes necessary to
move all electrons together, and for reasonable acceptance ratios
one must choose an increasingly smaller time step as the system
size increases. However, the more accurate nodal surface gives
both a quantitative improvement in properties and a qualitative
changes in some properties such as Fermi liquid
parameters\cite{kwon94}.

We also used the CEIMC\cite{dewing01} method to generate a
collection of proton positions appropriate to liquid metallic
hydrogen at 5000K, far above the melting temperature of the
lattice. Using these configurations we tested the accuracy of the
same trial functions described above.  See Table IV. The values
marked BF3-O are obtained minimizing local energy and variance for
1000 different equilibrium configurations. We compared to the
other ways of determining the backflow functions, either the
analytic formulas (see the Appendix) or optimized on the lattice.
While the optimized BF3 functions have a slightly lower energy in
some cases, this does not compensate for the difficultly and
reliability of performing the optimization. We find that the BF3
wavefunctions are about 20mH/atom lower in energy that from the
SJ-PW at the VMC level, and have a lower variance. This comparison
shows that disorder weakly affects the determined functions at
least in this experiment. This supports our belief that the BF3
wavefunction is ``transferable'' to a variety of protonic
configurations. In addition, we expect the backflow wavefunction
to be more effective in the disordered system, since the energy
degeneracies caused by crystal symmetry of a perfect lattice are
not present. Comparisons using optimized LDA functions to support
this hypothesis will be reported  in a future publication.

\begin{table}
\begin{tabular}{|c|c|c|}\hline

wavefunction& $E_v$ & $\sigma^2$\\\hline
 SJ-PW &-0.4225(8) & 0.0812(4)\\
 BF3-O-bcc ep& -0.4418(5)& 0.0447(7)\\
 BF3-O-liq ep&-0.4433(8)& 0.0710(10)\\
 BF3-O-liq ee+ep&-0.4462(8)& 0.0482(8)\\
 BF3-A ee+ep& -0.4430(4)& 0.0548(2)\\
 BF3-A ee+ep+b& -0.4464(6)& 0.052(2)\\\hline
\end{tabular}
\caption{ Energy and variance of liquid metallic hydrogen at
rs=1.31, and $N=16$. The notation of the trial function is
described in Table II.  The entries marked (bcc) are performed
with the value of the parameters optimized on the perfect bcc
lattice. The other entries are optimized over 1000 independent
protonic configurations taken at thermal equilibrium at 5000K. All
results are using VMC.}
\end{table}

\section{Conclusion}

What we have shown in this paper is that ideas from perturbation
theory can be used to generate an explicit trial wavefunction
beyond the pair level. This gives us both an insight into the form
of the many-body wavefunction and a more efficient quantum Monte
Carlo simulation for disordered systems. This approach has also
given intuition on the effect of an external potential on the
wavefunction, even for a single electron. We have shown that one can
approximate the band wavefunction (a  3d table of numbers for each
Bloch wave), with three 1D functions ($u, w,$ and $y$) valid for
all Bloch waves achieving reasonable accuracy. It should be
recalled that for the electron-proton system, there will be these
3 functions for the ee interaction and 3 functions for the ep
interaction. We have found analytical representations of these
functions accurate throughout most of the phase diagram of the
electron gas and promising for metallic hydrogen.

An important consideration in Monte Carlo is computational
efficiency. For electron-electron backflow, the code runs slower
due to having to move all the particles together. For
electron-proton backflow that is not the case.  You can still move
electrons one at a time since all the changes in the Slater matrix
are confined to a single column; each such matrix value is given
by a term of the same form as a classical force, allowing it to be
quickly computed once the ep distance have been computed.
Expansions of single body orbitals in a plane wave basis can be
quite time-consuming, especially when pseudopotentials are not
used.

But the most important advantage of the backflow wavefunctions is
that the form can be easily extended to put in effects of
electron-electron correlation on the nodes. The outstanding
problem in the simulation of quantum systems is the ``fermion sign
problem.'' If the nodal surfaces are accurately approximated, then
the ``fixed-node'' method will give accurate results.  The present
work, establishes new analytic properties of the backflow
functions and thus leads to important progress in understanding
nodal surfaces. In particular, the effect of long-range
interactions resulting in perturbations to the nodal surfaces is
important to establish. Strong short-range effects can be captured
either by energy minimization or by the nodal release algorithm,
which can solve for the exact wavefunction for relatively short
projection times or for small numbers of fermions.\cite{ca80}.
Fixing the relationship between the long-wavelength collective
coordinates and the nodal surfaces could be crucial in obtaining
accurate simulations for fermion systems.

In the above, we have discussed the use of backflow functions for
simple metals, using plane waves as the reference state. It is
straightforward to apply the approaches explored here to an
insulating state.  In that case, the reference state will be a
determinant of Wannier functions, in the simplest case, Gaussians.
The backflow ideas are applicable for suggesting improvements to
the resulting Slater-Jastrow function.  This will be considered in
future work. A related problem is how to treat bound states in
metallic liquid hydrogen in a more accurate way.

Backflow ideas are also useful at finite temperature. In that case
we need to know how density matrices will evolve going from high
temperature to low temperature\cite{dmc96}.  One knows how to put
in backflow at high temperature. The challenge is to smoothly
interpolate to zero temperature since it is clear that the
backflow potential must be a smooth function of temperature. In
the variational density matrix method\cite{vdm} one uses a
Hartree-Fock approach with a Gaussian basis to determined the
evolution of the nodal surface of the many body density matrix.
The various approaches we have described here in particular the
Bohm-Pines method, will be useful in understanding the temperature
dependence.

Another important problem is to generalize these methods to treat
electrons with core states. The formalism should generate good
trial functions in the valence region and can be used with either
all-electron methods, or pseudopotentials in that region. We hope
that with some modification the procedures we have discussed will
be useful in the core region as well.

The hospitality of M. Mareschal and CECAM is gratefully
acknowledged. This research was funded by NSF DMR01-04399, NASA,
DOE (CSAR) the CNRS-University of Illinois exchange agreement, the
CNRS, ENS-Lyon and the Dept. of Physics at the University of
Illinois. We acknowledge useful discussions with K. Schmidt, S.-W. Zhang, R.
M. Martin, R. Needs and G. Bachelet. Computational resources were
provided by CECAM, NCSA and CINECA (Italy) through the INFN
Parallel Computing Initiative.

\section*{ Appendix: Analytic expressions of the trial wavefunction}

In this section we summarize the analytic two-body, backflow and
polarization functions which describe the trial functions. We
start from the pair-product (Slater-Jastrow) wavefunction based on
the RPA approximation, using \be 2 n \tilde{u}_q^{ee}=-1 + \left(
1 + \frac{2 n \tilde{v}_q}{\varepsilon_q} \right)^{1/2}
\label{uee}\ee and \be 2 n \tilde{u}_q^{ep}= -\frac{2 n
\tilde{v}_q} {\varepsilon_q \left(  1 + 2 n
\tilde{v}_q/\varepsilon_q \right)^{1/2}} \label{uep} ,\ee where
$\varepsilon_q=\hbar^2 q^2/2m \equiv \lambda q^2$. Here $m$ is the
electron mass and $n$ is the electronic density. Using a trial
function with ee and ep Jastrow factors corresponds to the
following extended Hamiltonian, Eq.(\ref{BP11}), with:
\begin{eqnarray}
M_q^2=(\tilde{u}_q^{ee})^2 2 n \varepsilon_q
\\
M_q P_q= \tilde{u}_q^{ep}\tilde{u}_q^{ee} 2 n \varepsilon_q
\end{eqnarray}
Applying the unitary transformation
(\ref{S1}) to the wavefunction, it generates the backflow
potentials,
\begin{eqnarray}
y_{q}^{ee,\mathrm{int}}& = &
\frac{2 \lambda M_q^2}{\omega_p(q)(\omega_p(0)+\varepsilon_q)}
\\
y_{q}^{ep,\mathrm{int}}& = &
\frac{ 2 \lambda M_q P_q}{\omega_p(q)(\omega_p(0)+\varepsilon_q)}
\end{eqnarray}
where we used $\omega_p^2(q)=8 \pi \lambda e^2 n+
2.4 \, k_F^2 \lambda  \varepsilon_q + \varepsilon_q^2$ for the plasma
frequencies and $k_F$ is the Fermi vector. The screened
interaction between electrons, Eq.(\ref{eeint}), and between
electrons and protons, Eq.(\ref{epint}), can be treated by
perturbation theory. Summing up the particle-hole (bubble)
diagrams, using only the zero$^{th}$ order plane waves, leads to
coefficients $\alpha_{k_1,k_2,q}$ for the electron gas, as given
by Eq.(\ref{pert}), but with an effective interaction and
dielectric constant:
\begin{equation}
\tilde{v}_{\mathrm{eff}}^{ee}(q)=\tilde{v}_q-M_q^2, \quad
\epsilon_{\mathrm{eff}}(q,\omega)=
1-\tilde{v}_{\mathrm{eff}}^{ee}(q)D(q,\omega)
\label{veff}
\end{equation}
where and $D(q,\omega)$ is the real part of the Lindhard function.
As the pair product form already accounts for plasmons, we do not
consider any additional plasmon contributions to Eq.~(\ref{pert}).
Expanding Eq.~(\ref{pert}) around $k_i=0$, we obtain:
\begin{equation}
y_{q}^{ee,sr}  = \frac{[S(q)]^2}{2 q^2}
\frac{\tilde{v}_{\mathrm{eff}}^{ee}(q)}{\varepsilon_q
 \epsilon_{\mathrm{eff}}(q,\varepsilon_q)}
\end{equation}
to obtain this formula we have further approximated the sum over
occupied (unoccupied) states by $[S(q)]^2/4$ where $S(q)$ is the
ideal gas structure factor,
\begin{eqnarray}
S(q)=
 \left\{
 \begin{array} {l@{\quad : \quad}l}
\frac{1}{2}\left[
 \frac{3q}{2 k_F}-\left( \frac{q}{2 k_F}\right)^3 \right] &  q < 2k_F \\
 1 & q \ge 2 k_F.
\end{array}
\right.
\end{eqnarray}
The screened electron-proton interaction gives a similar term \be
y_{q}^{ep,sr} = \frac{2 }{\varepsilon_q q^2}
\frac{\tilde{v}_{\mathrm{eff}}^{ep}(q)}{\epsilon_{\mathrm{eff}}(q,0)}
\ee with the screened electron proton interaction
$\tilde{v}_{{eff}}^{ep}(k)=-(\tilde{v}_k-M_kP_k)$.

Adding these two contributions, the total backflow is: \be
y_{q}^{ee}= y_{q}^{ee,\mathrm{int}}+y_{q}^{ee,sr},
\label{fullbee} \ee
  \be y_{q}^{ep}=
y_{q}^{ep,\mathrm{int}}+y_{q}^{ep,sr}.
\label{fullbep} \ee

We also performed calculations with an additional ee Jastrow
function $\sim \tilde{v}_{\mathrm{eff}}^{ep}(q)/(\varepsilon_q
\epsilon_{\mathrm{eff}}(q,0)) $ but this form did not lower the
energy. Assuming this form disturbs the already correct limiting
behavior of the Jastrow part $u_q^{ee}$ and $u_q^{ep}$ for $q \to
0$ and $q \to \infty$, we took only the portion around the
logarithmic singularity  at $2k_F$, by using the following
additional Jastrow factor:
\begin{equation}
\tilde{u}_q^{ee,\mathrm{add}}=\frac{[S(q)]^2}{4\varepsilon_q}
\tilde{v}_{\mathrm{eff}}^{ee}(q) \left\{\frac{1}{\epsilon_{\mathrm{eff}}(q,\varepsilon_q)} -
\frac{1}{\epsilon_{\mathrm{eff}}(0,0)} \right\} \ee
\begin{equation}
\tilde{u}_q^{ep,\mathrm{add}}=\frac{1}{\varepsilon_q}
\tilde{v}_{\mathrm{eff}}^{ep}(q)
\left\{\frac{1}{\epsilon_{\mathrm{eff}}(q,0)} -
\frac{1}{\epsilon_{\mathrm{eff}}(0,0)} \right\} \ee We used
$\tilde{u}_q^{ep}+\tilde{u}_q^{ep,\mathrm{add}}$ for the total
electron-proton Jastrow potential, but only $\tilde{u}_q^{ee}$,
since the additional term $\tilde{u}_q^{ee,\mathrm{add}}$ did not
improve the variational energies of the electron gas. We used the
unsymmetrical form of the polarization with different left and
right components given by Eq.~(\ref{w1}) and Eq.~(\ref{w2}): \be
w_u^{ep}(r)=u^{ep,add}(r) \quad w_y^{ep}(r)=y^{ep}(r). \ee
Analogous forms were used for the electron-electron part.

For the case of metallic hydrogen we tried to take into account
the effects of a possible bound state on the electron-proton pair
and backflow potential. The single electron wavefunction $\phi_b$
considering only one bound state can be approximately written by
\be \phi_b \simeq \frac{A}{\sqrt{N}}\sum_i \varphi_b(|{\bf r}-{\bf
r}_i|) \ee where ${\bf r}_i$ is the position of the $i^{th}$
proton and the sum extends over all $N$ protons. As single
particle orbital we will take the hydrogen ground state,
$\varphi_b=(\pi a_b^3)^{-1/2} \exp(-r/a_b)$, with energy $e_b=1/2m
a_b^2$; $A\le1$ is a normalization taking into account the
non-zero overlap between orbitals on different sites. Using
Eq.~(\ref{fk}) we obtain for the scattering amplitude \be
f(e_b,{\bf p}) =-\sum_i \frac{A e^{-i {\bf p} \cdot {\bf r}_i}
}{\sqrt{N \pi a_0^3}} \frac{4 \pi e^2}{p^2+(a_0^{-1}+k_{TF})^2 }
\ee where we have taken a screened Coulomb interaction $v(r)=-e^2
e^{-k_{TF}r}/r$ with the Thomas-Fermi wavevector $k_{TF}^2=2
k_Fe^2/\pi \lambda$, and we have neglected overlap effects from
different sites. From Eq.~(\ref{ck}) we can finally derive the
corrections to the pair potential,
\begin{eqnarray} u_q^{ep,b} & = & -
\frac{A}{\sqrt{n \pi a_b^3}(1+(q/2k_{TF})^2)}
\nonumber \\
& &\times \frac{4 \pi
e^2}{[q^2+(a_b^{-1}+k_{TF})^2][e_b+e_q]}
\end{eqnarray} and
\begin{eqnarray}
y_q^{ep,b} & = &- \frac{ 8 \pi e^2A}{\sqrt{n \pi a_b^3}(1+(q/2k_{TF})^2)}
\nonumber \\
& & \times
\left(\frac{1}{[q^2+(a_b^{-1}+k_{TF})^2]^2[e_b+e_q]}
\right.
\nonumber \\
& & \left.
+\frac{\lambda}{[q^2+(a_b^{-1}+k_{TF})^2][e_b+e_q]^2} \right)
\end{eqnarray}
for the backflow potential. We have cut-off the short-range part
of the corrections by multiplying with $[1+(q/2k_{TF})^2]^{-1}$ in
order not to destroy the cusp conditions.  Since we expect
a higher energy for the
ground state of the screened Coulomb interaction than for the
pure Coulomb potential, we used $a_b \approx 2 a_0$ and $A \approx 1$
in the numerical calculations.

All potentials were split into a short range and long range
part\cite{Natoli95} in such a way as optimize the accuracy for a
given r-space and k-space cutoff. The short range function is
evaluated in real space and the long range part is then calculated
by summing over Fourier components. Figure~\ref{etaf} shows
numerical values of $\eta (r)$ for the 3D electron gas. Comparing
with the same figure of Kwon et al. where these functions were
numerically optimized, we see that the short ranged functions are
very similar for $r_s <10$ but different at larger $r_s$.
Figure~\ref{w2f} shows the three-body contribution of $i^{th}$
wavefunction. It is a rapidly increasing function of $r_s$ and is
somewhat narrower and more structured than the numerically
optimized form.

\begin{figure}
\centerline{\psfig{figure=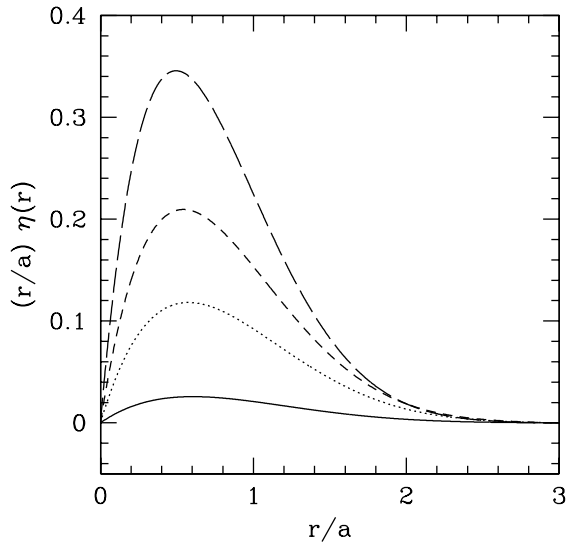,width=10cm,angle=0}} \caption{
The change in the quasiparticle coordinate $r \eta(r)$ (analytic
backflow) caused by an electron a distance $r$ away in the 3D
electron gas. Graphed is only the short range part of $\eta$ with
$N=54$. The four figures are for $r_s = 1,5,10,20$ from the bottom
to the top of the figure. Compare to the optimized forms in Fig.~(2)
in Kwon et al.\cite{kwon98}\label{etaf} }
\end{figure}

\begin{figure}
\centerline{\psfig{figure=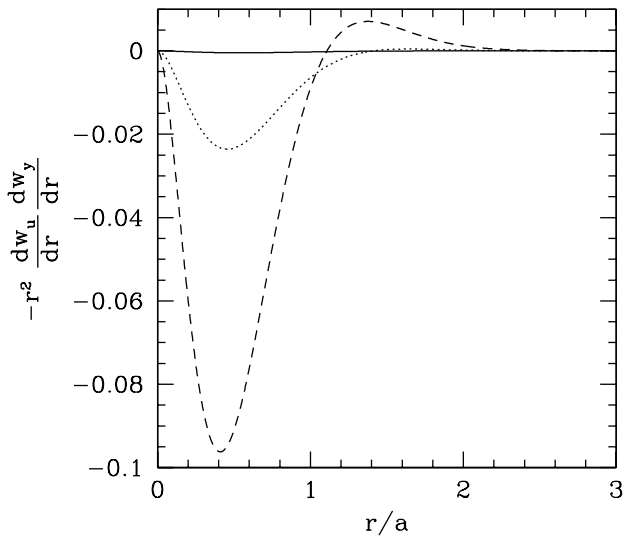,width=10cm,angle=0}} \caption{
The three-body contribution to the logarithm of the wavefunction
due to three electrons in the 3D electron gas. This is just the
short range parts of $w(r)$ for $N=54$. The solid line, for
$r_s=1$, is close to zero (maximum magnitude of $3 \times
10^{-4}$). The dotted line and dashed lines are for $r_s = 5,10$.
Compare to the optimized forms in Fig. (1) in Kwon et
al.\cite{kwon98}\label{w2f} }
\end{figure}



\begin{thebibliography}{10}
\bibitem{footnoteBT}
In principle, a parameterized backflow function could be used
within conventional band theory to avoid having to rediagonalize
the Hamiltonian away from the band center. For example one could
solve only for k=0 and use an analytic BF form to compute the band
structure at other values of k.
\end{thebibliography}
\end{document}